\documentclass[useAMS,usenatbib]{mn2e}

\def\mrc{MRC~2104$-$242~}
\def\spose#1{\hbox to 0pt{#1\hss}}
\def\simlt{\mathrel{\spose{\lower 3pt\hbox{$\mathchar"218$}}
     \raise 2.0pt\hbox{$\mathchar"13C$}}}
\def\simgt{\mathrel{\spose{\lower 3pt\hbox{$\mathchar"218$}}
     \raise 2.0pt\hbox{$\mathchar"13E$}}}

\author[Villar-Mart\'\i n et al.]{M. Villar-Mart\'\i n$^1$, S.F. S\'anchez$^2$, C. De Breuck$^3$, R. Peletier$^4$,
J. Vernet$^3$,  A. Rettura$^{3,5}$
\newauthor N. Seymour$^6$,  A. Humphrey$^7$,   D. Stern$^{8}$, S. di Serego Alighieri$^9$, R. Fosbury$^{10}$\\
$^{1}$Instituto de Astrof\'\i sica de Andaluc\'\i a (CSIC), Aptdo. 3004, 18080 Granada, Spain\\
$^{2}$Centro Astron\'omico Hispano Alem\'an de Calar Alto, E4004 Almer\'\i a, Spain \\
$^{3}$European Southern Observatory, Karl Schwarschild Str, 2, D-85748 Garching bei M\"unchen, Germany\\
$^{4}$Kapteyn Astronomical Institute, Univ. of Groningen, Postbus 800, 9700 AV  Groningen, The Netherlands\\
$^{5}$Universit\'e Paris-Sud 11, Rue Georges Clemenceau 15, Orsay, F-91405, France\\
$^{6}$Spitzer Science Center, California Institute of Technology, Mail Code 220-6,
1200 East California Boulevard
Pasadena, CA 91125 USA \\
$^{7}$Dept. of Physical Sciences, University of Hertfordshire, College Lane, Hatfield, Herts AL10 9AB, UK\\
$^{8}$Jet Propulsion Laboratory,California Institute of Technology, MS 169-506, Pasadena, CA 91109, USA\\
$^{9}$INAF-Osservatorio Astrofisico di Arcetri, Largo Enrico Fermi 5, I-50125 Firenze, Italy\\
$^{10}$ST-ECF, Karl-Schwarzschild Str. 2, 85748 Garching bei M\"unchen, Germany}
\title[]{VIMOS-VLT and Spitzer observations of a radio galaxy at $z=$2.5\thanks{Based on observations carried out at the
European Southern Observatory,  Chile (programs 073.B-0189, 065.P-0579 and 075.B-0729). and with the Spitzer Space
Telescope.}}

\begin{document}

\date{Accepted 2005 October 25.  Received 2005 October 25; in original form 2005 September 05}

\pagerange{\pageref{firstpage}--\pageref{lastpage}} \pubyear{2002}

\maketitle

\label{firstpage}

\begin{abstract}
We present: 1) a kinematic and morphological study of the
giant Ly$\alpha$ nebula associated with the radio galaxy MRC~2104$-$242
($z = 2.49$) based on integral field spectroscopic VIMOS data from VLT
; 2) a photometric study
of the host (proto?) galaxy   based on {\it Spitzer Space Telescope} data.

 The galaxy appears to be
embedded in a giant ($\ga 120$ kpc) gas reservoir that surrounds it
completely.  The kinematic properties of the nebula suggest that it is
a rotating structure, which would imply a lower limit to the dynamical
mass of $\sim3 \times 10^{11} M_\odot$.  An alternate scenario is that
the gas is infalling.  Such a process would be able to initiate and
sustain significant central starburst activity, although it is likely to contribute with less than 
10\% of the total stellar mass of the galaxy.

The near- to mid-IR spectral energy distribution of the radio galaxy
suggests the existence of a reddened, $E(B-V)=0.4 \pm0 .1$, evolved stellar
population of age $\ga 1.8$ Gyr and mass $(5 \pm 2) \times 10^{11}
M_\odot$.  The implied formation redshift is $z_f \ga 6$.  This
stellar mass is similar to the stellar masses found for massive
early-type galaxies at $z\sim$2 in deep, near-infrared surveys.

\end{abstract}


\section{Introduction}

In a study based on long-slit, medium resolution (R$\sim$700) Keck II/VLT
spectroscopy of a sample of 10 ultra steep spectrum, high redshift ($z\sim 2.5$) radio galaxies  (HzRGs),  we found that  these
objects are embedded in giant (often $\ge$100
kpc), low surface brightness   halos  of metal rich, ionized gas
with {\it quiescent kinematics} (Villar-Mart\'\i n et al.  2003 [VM03]). By quiescent we mean that the kinematics
 do not appear to be perturbed by  
interactions between the radio and optical structures.  The  halos usually extend well beyond the
brighter, perturbed regions and sometimes beyond the radio  structures.   
   We proposed that the quiescent halos
 are the  gas reservoirs from which the
galaxies started to form and might still be forming.
At some point, an active nucleus  switches on and the halo becomes observable 
thanks to strong line emission powered by the ionizing  continuum. 

Our work was seriously limited by the lack
of spatial information in directions other than the radio axis, through which the long slits were aligned. For this
reason, we are carrying out a programme of 3D integral field spectroscopy to characterize
the morphological, kinematic and ionization properties in two spatial
dimensions  of the extended 
 ionized gas in a sample of powerful radio galaxies at $z\sim$2-3. 

In this letter, we present the results obtained for the radio galaxy \mrc 
at $z=$2.49 (McCarthy et al. 1990).  The radio source  has an angular extent of 24$\arcsec$ ($\sim$200 kpc).
The  WFPC2 and NICMOS HST images  (Pentericci et al. 2001) show that the host galaxy is 
very clumpy.   A filamentary component 
 more than 2" long  is  found 
aligned with the radio axis to within a few degrees.   This region is likely
to be  undergoing a strong interaction between
the radio and optical structures (VM03, Humphrey
et al. 2005).

The existence of an evolved stellar population in HzRGs
has been subject of intensive study for the last two decades. The UV and optical restframe spectroscopic studies have only marginally detected the continuum, which may still contain significant non-stellar contribution (Vernet et al. 2001). To detect the old stellar
population we need to observe at rest-frame near-IR wavelengths. This is one of 
the scientific goals of the Cycle~1 {\it Spitzer} survey of 70 high redshift radio galaxies (P.I. Stern, program ID 3329).  We present
results for \mrc in this paper.

A  $\Omega_{\Lambda} =$ 0.73, $\Omega_{m}$ = 0.27 and $H_{0}$ = 71 km  $s^{-1}$  Mpc$^{-1}$ cosmology is adopted in this paper.

\section{Observations and data reduction}

The spectroscopic observations were made on   UT 2004 June  17 (ESO programme  073.B-0189(A)) using the
  VIsible MultiObject Spectrograph (VIMOS, Le F\'evre et al. 2003),  on the Nasmyth focus
of the UT3 VLT unit. The instrument  is equipped with an integral field unit  made
 of 1600  microlenses coupled to fibres, covering 27\arcsec x 27\arcsec on the sky with 0.67" diameter  fibres for the high spectral resolution mode.  The HR$_{\rm blue}$ grating was used, with  an effective wavelength range
 $\sim$4150-6200 \AA~  and an instrumental profile  3.0$\pm$0.5 \AA. The total integration time was 4.5 hr (9$\times$1800 sec).
The seeing FWHM during the observations was in the range $\sim$1.0-1.4$\arcsec$.

The data were reduced using a modified version of P3d 
(Becker 2002), in addition to our own software optimized for  VIMOS data (S\'anchez \& Cardiel 2005). The data were bias subtracted. 
The expected locations of the spectra were traced on a continuum-lamp exposure
obtained before each target exposure. 
The fibre-to-fibre response  at each  wavelength
 was determined from a continuum-lamp exposure. 
 Wavelength calibration was performed using  arc lamp spectra
 and
the telluric emission lines in the science data.  After
these basic reduction steps, a data cube was created for each exposure. The
cubes were then recentred spatially by determining the centroid of a nearby star at
each wavelength. This recentering corrects for
differential atmospheric refraction. The cubes were then combined using IRAF tasks, masking the broken and/or low sensitivity
fibres.  A 3$\sigma$ clipping algorithm removed cosmic
rays.
The sky background spectrum  was
estimated before subtraction  by averaging spectra of
object free areas of the data cube. The flux calibration was done using the long slit FORS-VLT
spectrum discussed in VM03.
Cross-talk effects were found to be negligible. These were evaluated using the field star in the VIMOS field of
view. The fibre to fibre contamination was found to be  $<$5\% for adjacent spectra in the CCD, dropping to less than 0.1\% for the 3rd adjacent spectra.

To  overlay the radio core position  (Carilli et al. 1997) on the Ly$\alpha$ image, we used the NICMOS HST image, where the radio core  was assumed
to coincide with the IR peak. This  is probably not the case if the nucleus
is heavily reddened. The 
 NICMOS peak position
  in the Ly$\alpha$ image was then calculated using a bright star that is
present in both images.    The  1$\sigma$ uncertainty
 in the Ly$\alpha$-radio
registration is $\sim$0.3".

 Details about the observations and data reduction for the {\it Spitzer} data will be presented by Seymour et al. (in prep.).
Briefly, for MRC~2104$-$242, we  obtained 3.6, 4.5, 5.8 and 8.0\,$\mu$m images on UT 2004 October 27 using 
the InfraRed Array Camera (Fazio et al. 2004), and a 16\,$\mu$m image in ``peak-up only mode'' 
on UT 2004 October 21 with the InfraRed Spectrograph (Houck et al. 2004). 
The IRAC ``basic calibrated data'' exposures were mosaiced using the {\tt MOPEX} software package (http://ssc.spitzer.caltech.edu/postbcd/) with the pixels being resampled by a factor of 2. 
 We used {\tt SEXTRACTOR} (Bertin \& Arnouts 1996) to measure the photometry in a matched 7\arcsec\ 
aperture defined by the 3.6\,$\mu$m image. To obtain total magnitudes, we applied aperture corrections from 
in flight point spread functions (Lacy et al. 2005).

\section[]{Results}

A Ly$\alpha$ image (Fig.~1) was created by adding all the frames in the spectral
direction of the data cube across the
Ly$\alpha$ emission line spectral profile, covering the $\sim$4225-4260 \AA\ spectral range. 
 The  large equivalent width of the line in all positions where it is
detected ensures 
 that the contamination by continuum emission is negligible everywhere
across the image. Fig.~1 shows  two spatial components (SW and NE) separated
by $\sim$6 arcsec as previously described  by 
 McCarthy et al. (1990).
The radio core is located  between
the two    clumps,  where  Ly$\alpha$ presents
a minimum. Similar Ly$\alpha$ morphologies have been
 observed in  other HzRGs (e.g. Reuland et al. 2003).
The SW component  overlaps partially with the filamentary structure described
in \S1.

\begin{figure}
\includegraphics{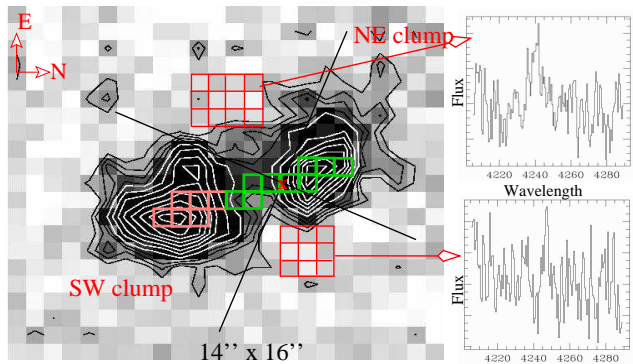}
\vspace{2.0in}
\caption{Ly$\alpha$ image of \mrc. Contours correspond to $\sim$5, 10, 15, 20, 25, 35, 45, 60, 70, 80, 90 and 100 \% of the
maximum contour value. Pixels marked in green  and pink  colours were
used to extract the spectra in Fig.~2.
Ionization cones with an opening angle of 90$\degr$
 and a vertex at  the radio core position (red  'x') are shown (black lines). 
 The spectra of  two
  intercone regions (red squares) are also shown. Ly$\alpha$ is detected in both cases.
}
\end{figure}
Low surface brightness Ly$\alpha$ emission is detected in regions that
seem to be
 outside any plausible ionization cones. Two spectra were extracted from the two intercone regions shown in Fig.1, where we
 have assumed a typical cone opening angle of
 90$\degr$  and the position of the
cone vertex coinciding with the radio core.   Ly$\alpha$, 
although noisy, is clearly detected in the two intercone regions.
Any rotation of the
 axis or reasonable shift of the  vertex position  leads to some of the
high surface brightness emission lying outside the ionization cones.
 McCarthy et al. (1990)  mentioned the detection of a  diffuse Ly$\alpha$ halo  that surrounds the entire object. 
 If the gas is ionized (but see Villar-Mart\'\i n, Binette \& Fosbury 1996),   the ionization mechanism is
an  open question.
 Hot young stars and cooling radiation are two interesting possibilities.

\subsection{Ly$\alpha$ kinematics}

Our goal is to map the 2D kinematic field of the quiescent gas
(i.e. non perturbed by jet-induced shocks)
 and use it as a source of information about its origin and
the   formation process of the galaxy. If the gas motions are gravitational, we
can  constrain the dynamical mass of the system. Because of the high sensitivity of
Ly$\alpha$ to absorption by HI, it is  often an unreliable kinematic tracer.
Given the low S/N of the data, 
we cannot  use other  reliable  tracers, such as the HeII$\lambda$1640 line.
 Instead,
we have
compared the integral field results here with  our earlier kinematic study (VM03) 
based on FORS2-VLT moderate resolution (6 \AA) 
long slit spectroscopy, which made use of both Ly$\alpha$ and HeII.

We show in Fig. 2 the 1-dimensional VIMOS spectra of  the SW and NE 
clumps
using two apertures (see Fig.1) matching as much as possible  apertures {\it 4} 
and {\it 3} of VM03.
Ly$\alpha$  in the SW clump is  split into two 
 components (blue (SWB)
and red (SWR)), both of which are well represented by Gaussian profiles.
The line 
 FWHM  are  470$\pm$20 and 550$\pm$20   km s$^{-1}$  
respectively. They are separated by 820$\pm$20 km s$^{-1}$.
 The line  in the NE component is  well represented
by a Gaussian profile of
  FWHM  630$\pm$20  km s$^{-1}$ redshifted  by 250$\pm$10  km s$^{-1}$
relative to the SWR component. 
These results are
in good agreement with those obtained with the FORS2-VLT data.  
In particular, the
HeII line in the VM03 long-slit spectrum of the SW clump
shows the same split as seen in the current VIMOS Ly$\alpha$ profile.
Since the HeII line is much less susceptible to absorption, this
implies that the double structure is due to kinematics rather than
absorption.

\begin{figure}
\includegraphics{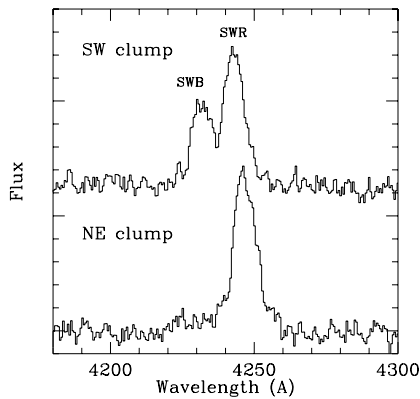}
\vspace{2.0in}
\caption{1-dimensional Ly$\alpha$ VIMOS spectra extracted from the NW and SE clumps, using the regions shown in Fig.1.  The flux levels  have been shifted for clarity.
The clear split observed in the SW  Ly$\alpha$ profile   is  due to 
kinematics. We argue that
the SWR  and  NE components  are both  quiescent gas.
}
\end{figure}

We argue that the red Ly$\alpha$ component from the SW clump (SWR) and the Ly$\alpha$ emission
from the NE clump are emitted by quiescent gas.
The remarkable similarity in the kinematic properties (see below) of
 these two components
strongly suggest that the gas in both regions follows a similar
kinematic pattern with the same origin. 
 On the other hand, the sharp
and large velocity shift of the SWB component suggests that this is a 
different, decoupled kinematic component.  
Although   detected almost all over the SW clump, this component is strongest relative to the  SWR component in  pixels
adjacent and coincident with the filamentary structure (\S1), where
strong evidence for jet-gas interactions exist (Humphrey et al. 2005), suggesting
that   it is emitted by perturbed gas.

We find that the Ly$\alpha$ profile of the quiescent gas is 
not strongly distorted by
absorption.
VM03 found that in general, although
 the kinematic
measurements (FWHM and velocity shifts) for Ly$\alpha$ and HeII  can be rather different for the perturbed  gas (usually broad lines), they are are in  good
agreement within the errors for the quiescent gas. This is also the case for \mrc (see fig.7, VM03).  Therefore, the
quiescent Ly$\alpha$ emission is less affected by absorption,
~probably due~ to the narrowness of the lines.  

In summary, the Ly$\alpha$ emission from the quiescent gas (components NE and
SWR) can be
isolated in \mrc at each spatial location and it can be used as   a reliable kinematic tracer.

\begin{figure}
\includegraphics{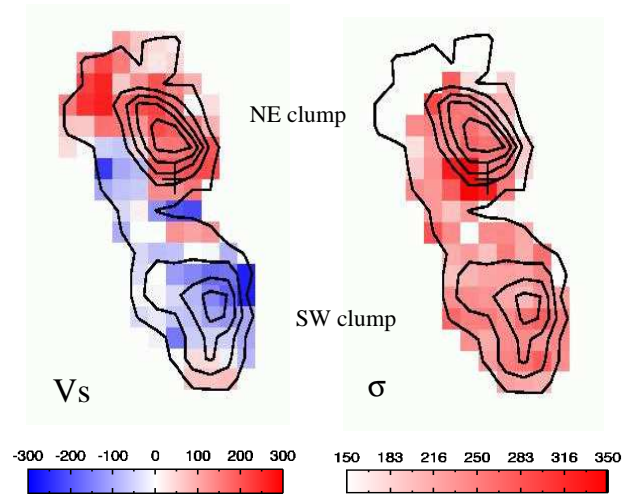}
\vspace{2.3in}
\caption{Overlay of the VIMOS Ly$\alpha$ intensity contours of the quiescent gas (i.e., components NE and SWR) with
the Ly$\alpha$  velocity field (left) and line width $\sigma$ (right).    
 The position of the  radio core is indicated with
a cross.  Zero velocity corresponds to a redshift $z= 2.490$.
The velocity field appears symmetric and ordered, suggesting either
 rotation or radial flows.}
\end{figure}

We now analyse
the velocity field of the {\it quiescent} gas. The Ly$\alpha$ profile was fit at each spatial
pixel with  Gaussian profiles.  The results are 
 shown in Fig.~3 (left) with the intensity contours of the quiescent gas overplotted. The errors
are $<$30 km s$^{-1}$ for most 
pixels.
The  velocity dispersion $\sigma$ map  (corrected for instrumental broadening) is also shown (Fig.~3, right).  
 The error is estimated to be $<$60 km s$^{-1}$. 

The kinematic pattern is rather symmetric and 
apparently ordered.
  The overall velocity field 
 is very suggestive of a 
rotating structure, for which the half amplitude of the rotation 
curve is $v_{rot}\sim$150 km s$^{-1}$. The $\sigma$ map is quite featureless 
with  $\sigma \sim$200-280 km s$^{-1}$ 
for most spatial positions, with little noticeable  difference between the two clumps other than a slight broadening of the line near the radio core. 
If the active nucleus coincides with the
 dynamical center of the system, this line broadening 
further supports rotation.

\subsection{Spitzer data}

Table~2 presents the {\it Spitzer} photometry for MRC~2104$-$242, as well as
near-infrared photometry from ground-based ($J_s$,$K_s$) ISAAC-VLT (ESO programme  075.B-0729(B)) and {\it
HST} (F160W) images.  We have corrected the   near infrared
photometry for emission line contamination using archival VLT/ISAAC $JHK$ spectroscopy (ESO programme 65.P-0579(A)). We determine the contributions to be 15$\pm$5\%, 76$\pm$5\%, and 51$\pm$5\% to the total flux in $J_s$, $H$ and $K_s$, respectively.

Fig.~4 plots the near- to mid-IR spectral energy distribution (SED).
The SED shows a clear rise between the $K-$ and IRAC 3.6\,$\mu$m
bands, and again between 8.0 and 16\,$\mu$m.  We now argue that the
3.6 to 5.8\,$\mu$m points are dominated by an old stellar population.
Figure~4 shows that the upturn in the SED between 8.0 and 16\,$\mu$m
(see Table 2) is unlikely to be  explained by a stellar population;
the 16\,$\mu$m point is most likely dominated by warm dust emission
associated with the active nucleus.  We assume the 16\,$\mu$m point
can be fit with a simple power law of slope $-3$ (Fig.~4), as determined
from a sample of 12 radio galaxies with {\it Spitzer} 16 and
24\,$\mu$m data (Seymour et al., in prep.).  Scaling this
power-law to the 16\,$\mu$m point implies that we can safely ignore
any thermal dust emission at $\lambda_{\rm obs} \leq 8.0\,\mu$m.
This is further supported by the 8.0\,$\mu$m upper limit.

Similarly, the $JHK$ points allow us to constrain the combined
contributions due to scattered quasar light, nebular continuum and
young stellar populations.  These contributions all have a blue
slope, while we detect a clear rise in the SED between 2.2 and
3.6\,$\mu$m. We can therefore constrain these combined contributions
 to $\ll$5\% in the IRAC bands. Note that the
excess of the $J$ band flux over the model predictions (Fig.~4) and
over the $H$ band measurement can only be explained by such
contributions, since  any old stellar population  would be fainter in
J than in H (especially when we invoke reddening).
 Hence our strongest constraint comes from the $H$
band detection.  Based
on restframe near-infrared spectroscopy of low-redshift
active galaxies (e.g. Sosa-Brito et al. 2001), we expect emission
line contributions to the IRAC bands to be $<$5\%.

We now use the $H$ thru 8.0\,$\mu$m SED of \mrc\, to derive the
most probable stellar mass of the host galaxy (see, e.g., Yan et al. 2005 for a detailed
discussion
on uncertainties).  The $J$-band point
is omitted as it appears to be contaminated by light associated
with either the buried AGN or a young stellar population, and the
16\,$\mu$m point is ignored as its likely non-stellar in origin.
We systematically compare the observed SED with a set of templates
computed with P\'EGASE.2 models (Fioc \& Rocca-Volmerange 1997) via
a $\chi^2$ minimization technique.  Star formation histories (SFHs),
ages, masses and dust extinction are free parameters.
We model simple stellar populations  of different metallicities
using a Salpeter (1955) IMF with a lower mass limit of $0.1 M_{\odot}$
and an upper mass limit of $120 M_{\odot}$.  We adopt SFHs that
have been shown to consistently reproduce observations of $z\sim
0$ passively-evolving, early-type galaxies (Fioc \& Rocca-Volmerange 1999).  A
more detailed description of the fitting method will be
presented in Rettura et al. (2005, in prep.).

\begin{table}
\centering
\begin{tabular}{llll}
\hline
     & $\lambda_{\rm cent}$ & FWHM & $f_\nu$ \\  
Band & ($\mu$m)  & ($\mu$m) & ($\mu$Jy)  \\ \hline
$J_s$  &    1.2   &   0.16  & 4.1$\pm$1.5 \\
F160W  &     1.6 &    0.43 &  1.7$\pm$1.2 \\ 
$K_s$   &    2.2  &  0.27 &  4.8$\pm$1.8 \\  
3.6$\mu$m  &    3.6  &  0.7 &  27$\pm$3 \\  
4.5$\mu$m &    4.5  &  0.99 &  29$\pm$3 \\ 
5.8$\mu$m &   5.7   & 0.90 &  32$\pm$7 \\ 
8.0$\mu$m  &  8.0    & 2.8 &  $\leq$35  \\    
16.0$\mu$m &   16.0  & 6.0 &  72$\pm$18  \\ 
\hline
\end{tabular}
\caption{Near to mid-IR photometry of \mrc.  The $J_sHK_s$ photometry has
been corrected for line emission (see text). }
\end{table}

\begin{figure}
\includegraphics{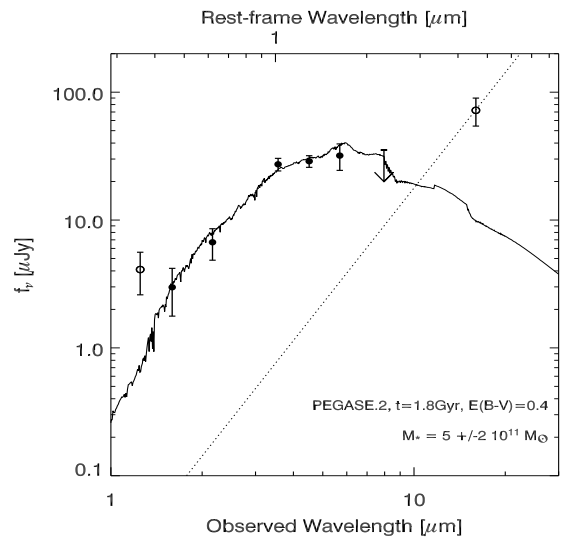}
\vspace{2.6in}

\caption{Near to mid-IR SED of \mrc.  The best fit is produced by
a reddened, relatively evolved population of age $1.8-2.5$ Gyr and
stellar mass $(5 \pm 2) \times 10^{11} M_{\odot}$.  The $J_s$ band
point is dominated by a combination of scattered quasar, nebular
continuum and young stellar emission. All of these contributions
have a blue slope, and will therefore contribute $\ll$5\% to the
IRAC photometry. A power law of index -3 (dotted line) represents
the expected thermal dust emission. The figure shows that its 
contribution can be 
safely ignored 
below 8 $\mu$m.}

\end{figure}

Results of the best-fit model  are shown in Fig.~4.  We estimate
the errors for the mass by sampling the full probability distribution
in the multidimensional parameter space.  We find a best-fit stellar
mass estimate  of $(5 \pm 2) \times 10^{11} M\odot$ and a color
excess $E(B-V)=0.4 \pm 0.1$. This is  similar to the stellar masses found for the most massive, 
early-type galaxies at $z\sim$2 in deep, infrared surveys (e.g. Daddi et al. 2004,
Labb\'e et al. 2005).  The best-fit age is 2.5 Gyr, which is
very close to the age of the universe at the source redshift, (2.7
Gyr).  The minimum age that still provides a reasonable ($3 \sigma$)
fit is 1.8 Gyr, implying a formation redshift $z_f \simgt 6$.
Finally, we note that the SED cannot be fit by any simple or composite
stellar population without invoking substantial reddening.

\section{Discussion and Conclusions}

Let us assume that the giant Ly$\alpha$ nebula in \mrc\, is supported
by rotation.  If its morphology maps the true gas distribution and
the gas is settled on a disk, we derive an inclination angle $i
\sim 75 \pm 5\degr$ relative to the plane of the sky and a dynamical
mass $\sim (3.2 \pm 0.7) \times 10^{11} M_{\odot}$ within a radius
of 7 arcsec ($\sim 60$ kpc).  This is similar to the stellar mass
inferred from the {\it Spitzer} data (see \S3.2), suggesting a
surprisingly small dark matter fraction.  However, the derived
dynamical mass is likely to be a lower limit, since the gas is
expected to be anisotropically illuminated by the hidden quasar.
Therefore, the disk inclination and the direction of the major axis
cannot be determined (e.g., the gas we see might be close to the
minor axis).

If the giant halo in \mrc\, is settled on to a well defined disk,
then $\frac{\sigma}{v_{rot}}<1$, such that $v_{\rm rot}$ must be
$\simgt 240$ km s$^{-1}$ and $M_{\rm dyn} \simgt 8 \times 10^{11}
M_{\odot}$.  On the other hand, it takes several rotation periods
for gas to settle into a disk. If this is the case for the giant
nebula in \mrc\, and we assume a minimum of three rotation periods
($P_{\rm rot} = 2 \pi r/v_{\rm rot}$) for a well-ordered disk, the
2.7 Gyr age of the universe at $z = 2.49$ requires $v_{\rm rot} >$
400 km s$^{-1}$ for which $M_{\rm dyn} > 2 \times 10^{12} M_\odot$.
Smaller $v_{\rm rot}$ would imply that the gas has not yet settled
into a stable structure.

It is possible that the SW and NE clumps are two different objects
rotating around a common center (e.g. De Breuck et al. 2005).
However the fact that Ly$\alpha$ emission surrounds the object
completely suggests that it is a single, rotating structure.  If
the gas is rotating, this means that giant ($r > 60$ kpc) gaseous,
rotating structures (disks?) can exist already at $z \sim 2.5$.
These could be the progenitors of the giant ($>$several tens of
kpc) rotating disks discovered around several low redshift radio
galaxies and early-type galaxies (e.g. Morganti et al. 2002).  High
redshift cooling flow models for galaxy formation (e.g. Haiman,  
Spaans \& Quataert 2000)  predict the formation of a giant, rotationally-supported
disk as part of the process in less than 1 Gyr.  Therefore,
 the evolved ($\simgt 1.8$ Gyr old), stellar population revealed by the
{\it Spitzer} data  probably  formed well before the disk.

Alternatively, the general pattern of the  velocity field is also
suggestive of radial motions.  The radio properties
can discriminate between inflow and outflow.   The NE
radio lobe is brighter and more polarized (Pentericci et al.  2000). 
According to the Laing-Garrington effect (Laing 1988), this lobe
is closer to the observer.  Since the gas located at this side
of the nucleus is the most redshifted, it must be infalling.  The
collapse of such a giant structure  suggests that we
are witnessing an early stage in the formation process of the central galaxy
(e.g. Haiman, Spaans \& Quataert 2000).

Let us assume an infall velocity of $\sim 150 \times \sin \alpha$
km s$^{-1}$ and a radius $\sim 60 \times \cos \alpha$ kpc, where
$\alpha$ is the angle between the ionized cone axis and the line
of sight.  Since this is a narrow-line radio galaxy, $\alpha$ is
expected to be large, $\alpha \simgt 60 \degr$.  Therefore, after
$\simlt 2.5 \times 10 ^8$ yr, the 60 kpc halo should  collapse 
and disappear in a very small redshift interval (by $z
\sim 2.2$). The existence of giant, quiescent halos in radio galaxies
at very different redshifts (van Ojik et al. 1996, Villar-Mart\'\i n
et al. 2005) suggests that they are long-lived structures, unless
there are multiple episodes of Ly$\alpha$ halos.  The implication
is that much larger reservoirs of invisible gas must exist and/or
some mechanism must deccelerate the infall process at some point
(Iono et al. 2004). 

As in VM03 and using the geometrical information provided by the
VIMOS Ly$\alpha$ image, we estimate an ionized gas density $n \sim
55$ cm$^{-3}$ (see VM03 for uncertainties on this calculation) and
a gaseous mass (neutral and ionized) within 60 kpc of $\sim 10^{10}
M_{\odot}$.  A mass deposition rate in the range $\sim (40-250)
M_{\odot}$ yr$^{-1}$ is inferred.   This infall rate might be able
to initiate and sustain central starburst activity at several tens
of $M_{\odot}$ yr$^{-1}$ (Iono et al. 2004).  Since the object contains a $\geq 1.8$
Gyr old population as well as the young population triggered by the
current inflow, at least two episodes of star formation are inferred
for the formation of this massive galaxy. If the halo has $\sim$10$^{10}$ $M_{\odot}$, 
this is probably the maximum it will contribute to build up 
the stellar mass of the galaxy, so that the infalling gas will only
be $<$10\% of the galaxy mass.

In summary, \mrc\, contains an evolved stellar population that
formed at $z \simgt 6$. The kinematic properties of the extended
gas  suggest that it is associated with a giant
rotating structure, the remnant of the  formation process of the galaxy.  Alternatively, the gas might be infalling during an early phase in the galaxy
formation process.
Extending a similar study to a larger sample of objects will be
very useful to understand the demographics and nature of ordered
kinematics in the quiescent gas associated with HzRGs.

\section*{Acknowledgments}
We gratefully acknowledge the SHzRG team who have provided the {\it
Spitzer} data presented here. The work of MVM has been supported by the Spanish Ministerio de Educaci\'on y Ciencia 
and the Junta de Andaluc\'\i a through the grants AYA2004-02703 and TIC-114
respectively.  The work of DS was carried out at
Jet Propulsion Laboratory, California Institute of Technology, under
a contract with NASA.

\end{document}